\documentclass{iopconfser}
\usepackage{amsfonts}
\usepackage{amsmath}
\usepackage{braket}
\usepackage{booktabs}
\usepackage{graphicx}
\usepackage{makecell}
\usepackage[merge,numbers,sort,compress]{natbib}
\usepackage{subcaption}
\usepackage{hyperref}

\begin{document}

\title{Quantum algorithms for the simulation of QCD processes in the perturbative regime}

\author{
Herschel A. Chawdhry$^{1}$ and Mathieu Pellen$^{2}$}

\affil{$^1$Department of Physics, Florida State University, 77 Chieftan Way, Tallahassee FL, USA}
\affil{$^2$Albert-Ludwigs-Universität Freiburg, Physikalisches Institut, Freiburg, Germany}

\email{hchawdhry@fsu.edu, mathieu.pellen@physik.uni-freiburg.de}

\begin{abstract}
Theoretical predictions for high-energy collision processes at particle colliders, such as the Large Hadron Collider (LHC), rely on calculations in perturbative Quantum Chromodynamics (QCD), which are often computationally challenging.
In these conference proceedings, we explore the possibility of using quantum computers to simulate QCD processes in the perturbative QCD regime.
In particular, as a first step towards that goal, we present quantum circuits to simulate the colour part of perturbative QCD.
The circuits are validated by implementing them on a simulated quantum computer and verifying the colour factors for several example Feynman diagrams.
\end{abstract}

\section{Introduction}
Particle physics studies the smallest and most fundamental particles that constitute the universe.
A large portion of our knowledge in this field comes from studying particle collisions at high-energy colliders, such as the Large Hadron Collider (LHC).
One of the major components of the LHC research programme is the high-precision measurement of production rates for a wide variety of processes.
To best exploit these high-precision measurements, it is vital that theoretical predictions of comparably high precision are produced.
Comparing these theoretical predictions against the experimental measurements from the LHC can validate the Standard Model of Particle Physics, improve our understanding of subatomic physics, and potentially provide early clues about new particles or phenomena beyond the Standard Model.

Since the LHC collides hadrons (usually protons), theoretical predictions for LHC processes must employ Quantum Chromodynamics (QCD), the theory of the strong force governing the interactions of hadrons.
One obvious approach to QCD calculations would be to directly simulate the quantum fields on a space-time lattice.
It was mentioned in another talk at this conference that modern state-of-the-art multi-petaflop supercomputer-based lattice QCD simulations implement around $96-192$ lattice sites in each of the 4 space-time directions.
Since the LHC probes energy scales that are $\mathcal{O}(10^3-10^4)$ times larger than the QCD scale $\Lambda_{\textrm{QCD}} \sim 0.2 \textrm{ GeV}$, lattice-based predictions at the highest LHC energies would require thousands of lattice sites in each of the 4 space-time direction.
This is likely to remain infeasible for the foreseeable future, as the computational cost of lattice-based simulations scales as a large power (typically at least 6 or 7) of the lattice spacing.

Fortunately, at these high energies the QCD coupling parameter $\alpha_s$ becomes small and so it becomes possible to perform perturbative calculations, whereby instead of simulating quantum fields on a lattice, theoretical predictions for observables (such as particle production rates) are directly calculated in the form of perturbative series in $\alpha_s$.
These perturbative calculations have been widely successful in predicting the production rates and kinematics of a broad range of high-energy processes at the LHC.
The precision and accuracy of such predictions generally improves with the perturbative order in $\alpha_s$, but so does the computational complexity: indeed past experience shows a typical elapse of 1-2 decades to add one new order in $\alpha_s$ to the prediction of a given observable.
These computational obstacles present a key limiting factor in the precision of theoretical predictions for colliders, and so new calculational techniques and technologies are constantly being sought.

%Perturbative Quantum Chromodynamics (QCD) calculations provide high-precision predictions of the scattering of fundamental particles, especially at hadron colliders, and are therefore a vital part of the Large Hadron Collider (LHC) physics program.
%Calculational complexity presents a key limiting factor in producing these predictions and so the development of new computational techniques is central to advancing the state of the art.

In this conference proceedings paper, based on our article~\cite{Chawdhry:2023jks} and related proceedings~\cite{Chawdhry:2023ken}, we explore whether quantum computers in the future could help perform perturbative QCD calculations.
In particular, as a first step towards this goal, we focus on using quantum computers to simulate the colour part of perturbative QCD.

Quantum computers~\cite{Benioff:1979ce,Feynman:1981tf}, which operate by manipulating the wavefunctions of quantum-mechanical systems, are of interest because they are anticipated to provide a better cost scaling than classical computers, for certain problems.
Although examples of proposed applications exhibiting speed-ups are limited in number at present, they include impressive examples like an exponential speed-up for prime factorisation~\cite{Shor.365700} and quadratic speed-ups for so-called ``unstructured search'' problems~\cite{10.1145/237814.237866}, the latter class encompassing many computational problems that are otherwise solved by brute force (ranging from Monte Carlo integration to the mining of cryptocurrencies).
Another proposed application is the use of one quantum computer to simulate another quantum system, such as in computational quantum chemistry~\cite{ChemistryRev,RevModPhys92015003}, condensed matter physics~\cite{RevModPhys.86.153,qute.201900052}, and lattice QCD~\cite{Klco:2021lap,Bauer:2022hpo}.

%Quantum computing was first proposed 4 decades ago~\cite{Benioff:1979ce,Feynman:1981tf} and has been of great interest over the years because for certain problems it promises large speed-ups.
%In particular, it promises exponential speed-ups for prime factorisation~\cite{Shor.365700} and quadratic speed-ups for generic unstructured search problems~\cite{10.1145/237814.237866} (of which Monte Carlo integration is an example).
%A further application is the simulation of quantum systems: since quantum computers perform calculations by manipulating the quantum states of a system, it is natural to use a quantum computer to simulate other quantum systems.
%In particular, active fields of research exist studying methods to use quantum computers to perform simulations of quantum chemistry~\cite{ChemistryRev,RevModPhys92015003}, condensed matter systems~\cite{RevModPhys.86.153,qute.201900052}, and lattice QCD~\cite{Klco:2021lap,Bauer:2022hpo}.

Although recent years have seen many proposals (reviewed in Refs~\cite{Klco:2021lap,Bauer:2022hpo}) for using quantum computers to simulate lattice QCD, the possibility of performing quantum simulations of perturbative QCD has remained largely unexplored, with the exception of some work on parton showers~\cite{Bepari:2020xqi,Bauer:2019qxa,Bepari:2021kwv,Gustafson:2022dsq}.
Part of the reason for this may be that the aims of perturbative QCD are different from those of most quantum simulations: the latter (including lattice QCD) seek to calculate the time evolution of a known Hamiltonian, whereas perturbative QCD calculations seek to calculate the (Hermitian but not necessarily unitary) transition matrices describing the scattering of pre-determined initial and final states.

The aim of the work reported in this proceedings paper (and first reported in our article~\cite{Chawdhry:2023jks} to which we refer the reader for more comprehensive explanations) is to take the first steps towards the quantum simulation of generic perturbative QCD processes, by presenting quantum computing circuits to simulate the colour part of perturbative QCD calculations. The colour part was chosen as a useful starting point because it involves smaller Hilbert spaces than the kinematic parts, and therefore provides a good simplified setup with which to develop generic techniques and strategies that can in the future be built upon to implement the kinematic parts of the calculation.

There are several specific motivations for seeking to use quantum computers to simulate perturbative QCD. One motivation is that perturbative QCD requires the quantum-coherent combination of contributions from many unobservable intermediate states, making it a natural candidate to exploit the ability of a quantum computer to manipulate superpositions of quantum states.
Processes with high-multiplicity final states in particular could benefit from such calculations.
Another motivation is investigating the possibility to improve the speed or precision of perturbative QCD predictions by exploiting known quantum algorithms such as quantum amplitude estimation~\cite{Brassard:2000,Grinko:2019,Suzuki:2019,Nakaji:2020} and quantum Monte Carlo integration (see Ref.~\cite{Agliardi:2022ghn} and references therein).

Research on this topic is timely. After several decades of incremental progress in hardware and algorithms, recent years have seen companies such as IBM, Google, and Microsoft making large investments, producing hardware with up to a few hundred qubits (albeit prone to noise and lacking full connectivity between qubits), and aspiring to develop larger error-corrected quantum computers over a timespan of around a decade.
For this reason, many applications have been proposed in various areas of high-energy physics~\cite{Agliardi:2022ghn, Bauer:2019qxa, Bauer:2021gup, Bepari:2020xqi, Bepari:2021kwv, Bravo-Prieto:2021ehz, Cervera-Lierta:2017tdt, Chawdhry:2023jks, Clemente:2022nll, Cruz-Martinez:2023vgs, Fedida:2022izl, Gustafson:2022dsq, Kiss:2022pjw, Li:2021kcs, Perez-Salinas:2020nem, Ramirez-Uribe:2021ubp, Rigobello:2023ype, Williams:2023muq,Nicotra:2023rmn,Nagano:2023kge,Turco:2023rmx,Bass:2023hoi,DiMeglio:2023nsa,Bermot:2023kvh,Sborlini:2023uyq,Humble:2022klb,Hayata:2023bgh,Sborlini:2023mws,Brown:2023llg,Rodrigo:2024say,Ramirez-Uribe:2024wua,Galvez-Viruet:2024hry}.

\section{Quantum circuits for colour}\label{sec:circuits_for_colour}

In this section we will present quantum circuits that simulate the colour part of perturbative QCD.
The circuits are based on qubits, i.e.\ two-state quantum systems (such as spin half particles), which are represented pictorially by a quantum circuit diagram in which each qubit is drawn as a horizontal line.
The qubits are manipulated by performing operations, which are called gates by analogy to the \textsc{and} and \textsc{or} gates of classical computing.
A gate acting on $n$ qubits is defined by a $2^n$-by-$2^n$ matrix acting on the $2^n$-dimensional state space of those qubits. Since quantum mechanical operations are unitary, these matrices must be unitary.

Perturbative QCD calculations are often performed with the help of Feynman diagrams, with each diagram representing a contribution to the scattering amplitude (which is related to the transition matrix) for a given process. Given a Feynman diagram, the corresponding term in the scattering amplitude contains a factor $T^a_{ij}$ for each quark-gluon vertex, and a factor $f^{abc}$ for each triple-gluon vertex, where $T^a_{ij}$ are the generators of the Lie algebra $\mathfrak{su}(3)$ in the defining representation and $f^{abc}$ are the structure constants of $\mathfrak{su}(3)$.
Besides these colour factors, kinematic factors would normally also be required but they are neglected in this work as mentioned above.
As an example, the colour factor of the Feynman diagram shown on the left of Fig.~\ref{fig:quarkselfenergy} is
\begin{equation}\label{eq:self_energy_colour_factor}
\mathcal{C} = \sum_{
\substack{
a \in \{1, ..., 8\}\\
i,j,k \in \{1,2,3\}
}
} T^a_{ij} T^a_{jk} \delta_{ik},
\end{equation}
where the Feynman rules for the diagram require us to sum over intermediate states $j \in \{1,2,3\}$ and $a \in \{1, \ldots, 8\}$, and in this case we have further opted to trace over the initial colour $i$ and final colour $k$ of the quark line.

The generators $T^a_{ij}$ are linear operators, which are conventionally written in terms of the well-known Gell-Mann matrices $\lambda^a$ by defining $T^a = \frac{1}{2} \lambda^a$. Recalling that quantum gates are required to be linear operators, it is natural to ask whether the generators $T^a_{ij}$ can be implemented as quantum gates and thus be used to simulate the colour part of quark-gluon interactions.
As we will see, this can indeed be achieved although there are some obstacles.
One is that the Gell-Mann matrices are not of the form $2^n$-by-$2^n$, which would be desirable for the reasons described at the start of this section. 
Another obstacle is that Gell-Mann matrices are not unitary, but are instead Hermitian.
Our main article~\cite{Chawdhry:2023jks} includes explanations of how these issues are resolved.

The key results of this work are two quantum gates, $G$ and $Q$, which simulate, respectively, the colour parts of the triple-gluon interaction and the quark-gluon interaction in perturbative QCD.
By combining several of these gates, the colour parts of entire Feynman diagrams can be simulated.
This conference proceedings paper will only give an illustrative example of how these gates can be assembled together, and we refer the interested reader to our main article~\cite{Chawdhry:2023jks} for more comprehensive explanations as well as details about the explicit construction of the $G$ and $Q$ gates.

A gluon has 8 basis colour states, which we represent using the $2^3=8$ basis states of a register of 3 qubits.
A quark has 3 basis colour states, which we represent using 3 of the $2^2=4$ basis states of a pair of qubits, while the 4th basis state remains unused.

The $Q$ gate acts on a 3-qubit register representing a gluon and a 2-qubit register representing a quark, as well as a register $\mathcal{U}$ containing a few extra qubits whose purpose will be described below.
The $Q$ gate is designed such that if the gluon register is in a basis state $\ket{a}_g$, where $a \in \{1,\ldots,8\}$, and the quark register is in a basis state $\ket{k}_q$, where $k \in \{1,2,3\}$, and the $\mathcal{U}$ register is in a special reference state $\ket{\Omega}_{\mathcal{U}}$, then
\begin{equation}\label{eq:Q_gate_behaviour}
Q\ket{a}_g\ket{k}_q\ket{\Omega}_\mathcal{U} = \sum_{j=1}^3 T^a_{jk} \ket{a}_g\ket{j}_q\ket{\Omega}_\mathcal{U} + \left(\textrm{terms orthogonal to } \ket{\Omega}_\mathcal{U} \right) .
\end{equation}
If any of the registers are in superpositions of colour basis states or are entangled with other registers in the circuit, the $Q$ gate acts linearly on each element of the wavefunction, since quantum gates are linear.
The $Q$ gate can be seen to implement the Feynman rule for a quark line that emits or absorbs a gluon.
Note that while a circuit will in general contain several quark and gluon registers, it only contains a single $\mathcal{U}$ register. The purpose of the latter is to allow unitary gates like $Q$ to be constructed to implement, in the sense of eq.~\eqref{eq:Q_gate_behaviour}, linear but non-unitary operators like $T^a_{ij}$.
We therefore refer to $\mathcal{U}$ as the \emph{unitarisation register}.
Note that the size of $\mathcal{U}$ is small: it is logarithmic in the number of vertices in the Feynman diagram.

Similarly to the $Q$ gate, the triple-gluon gate $G$ is designed to act on any 3 gluon registers $g_1$, $g_2$, $g_3$ (each composed of 3 qubits as explained above) in colour basis states $\ket{a}_{g_1}$, $\ket{b}_{g_2}$, $\ket{c}_{g_3}$ in the following way:
\begin{equation}\label{eq:G_gate_behaviour}
G \ket{a}_{g_1}\ket{b}_{g_2}\ket{c}_{g_3}\ket{\Omega}_{\mathcal{U}} = f^{abc} \ket{a}_{g_1}\ket{b}_{g_2}\ket{c}_{g_3}\ket{\Omega}_{\mathcal{U}} + \left(\textrm{terms orthogonal to } \ket{\Omega}_\mathcal{U} \right) ,
\end{equation}
where $\mathcal{U}$ is again the same register which appeared in eq.~\eqref{eq:Q_gate_behaviour} and $\ket{\Omega}_\mathcal{U}$ is the same special reference state as before.
Equations~\eqref{eq:Q_gate_behaviour} and~\eqref{eq:G_gate_behaviour} can be interpreted to mean that when projected onto the special reference state $\ket{\Omega}_\mathcal{U}$ of the unitarisation register, the $Q$ and $G$ gates simulate the colour parts of a quark-gluon interaction and a triple-gluon interaction, respectively.

\section{Illustrative example}\label{sec:illustrative_example}
\begin{figure}[]
\center
        \begin{subfigure}{0.35\textwidth}
                 \includegraphics[width=\textwidth]{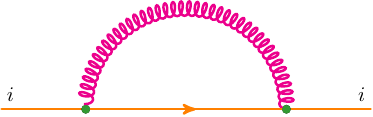}
        \end{subfigure}
\hfill
        \begin{subfigure}{0.60\textwidth}
                 \includegraphics[width=\textwidth]{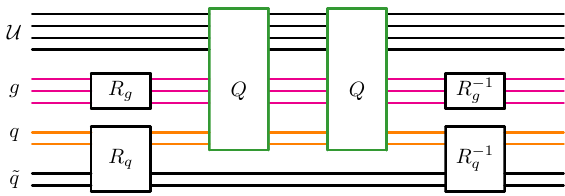}
        \end{subfigure}
        \caption{\label{fig:quarkselfenergy}%
                Example Feynman diagram (left) and a graphical representation of its corresponding circuit (right).}
\end{figure}
In this section we present an example to illustrate how the gates from section~\ref{sec:circuits_for_colour}, in particular $Q$, can be used to simulate the colour factor for a simple Feyman diagram.
A generalisation to arbitrarily complicated diagrams will be given in sec.~\ref{sec:generalisation}.
Let us consider the Feynman diagram which is shown in Fig.~\ref{fig:quarkselfenergy}.
The diagram has one gluon and one quark.
As was mentioned above, the colour of a gluon is represented by 3 qubits and the colour of a quark is represented by 2 qubits. However, there is a complication: in order to be able to compute the trace, we always introduce for each quark line a pair of 2-qubit registers $q$ and $\tilde{q}$, instead of just a single 2-qubit register.
The extra register $\tilde{q}$ is not affected by the simulation gates $Q$ or $G$; the only purpose for $\tilde{q}$ is to enable the trace to be computed, as we will see.

Initially, the circuit is in a reference state $\ket{\Omega}_g\ket{\Omega}_q\ket{\Omega}_{\tilde{q}}\ket{\Omega}_\mathcal{U}$, where $\ket{\Omega}_r$ indicates that each qubit of a register $r$ is in the state $\ket{0}$.
Next, a gate $R_g$ is applied to the gluon register to place it in an equal superposition of colour basis states:
\begin{equation}
\label{eq:Rg_effect}
R_g\ket{\Omega}_g = \sum_{a=1}^8 \frac{1}{\sqrt{8}} \ket{a}_g.
\end{equation}
The explicit form of $R_g$ is omitted here but it can be found in the Appendix of our main article~\cite{Chawdhry:2023jks}.
Next, the gate $R_q$ (which is also defined in the Appendix of Ref.~\cite{Chawdhry:2023jks}) is applied to the pair of quark registers as follows:
\begin{equation}\label{eq:Rq_effect}
R_q \ket{\Omega}_q \ket{\Omega}_{\tilde{q}} = \sum_{k=1}^3 \frac{1}{\sqrt{3}} \ket{k}_q\ket{k}_{\tilde{q}}.
\end{equation}
Here it can be observed that the $q$ and $\tilde{q}$ registers are entangled.
Thus, after applying the $R_g$ and $R_q$ gates, the quantum computer is in the following state:
\begin{equation}
\frac{1}{\sqrt{24}} \sum_{a=1}^8 \sum_{k=1}^3 \ket{a}_g\ket{k}_q\ket{k}_{\tilde{q}}\ket{\Omega}_\mathcal{U}.
\end{equation}

We will now perform the key simulation steps, by applying two $Q$ gates which correspond directly to the two interaction vertices shown in the Feynman diagram in Fig.~\ref{fig:quarkselfenergy}.
As already mentioned, the $Q$ never acts on the $\tilde{q}$ register.
It can be seen with the help of eq.~\eqref{eq:Q_gate_behaviour} that after applying the first $Q$ gate, the state of the quantum computer becomes
\begin{equation}
\frac{1}{\sqrt{24}} \sum_{
\substack{
a \in \{1,\ldots,8\}\\
j,k \in \{1,2,3\}
}
} T^a_{jk} \ket{a}_g\ket{j}_q\ket{k}_{\tilde{q}}\ket{\Omega}_\mathcal{U} + \left(\textrm{terms orthogonal to } \ket{\Omega}_\mathcal{U} \right)
\end{equation}
and after applying a second $Q$ gate, the state becomes
\begin{equation}\label{eq:example_circuit_after_second_Q_gate}
\frac{1}{\sqrt{24}} \sum_{
\substack{
a \in \{1,\ldots,8\}\\
i,j,k \in \{1,2,3\}
}
} T^a_{ij}T^a_{jk} \ket{a}_g\ket{i}_q\ket{k}_{\tilde{q}}\ket{\Omega}_\mathcal{U} + \left(\textrm{terms orthogonal to } \ket{\Omega}_\mathcal{U} \right) .
\end{equation}
While this bears a resemblance to the desired colour factor in eq.~\eqref{eq:self_energy_colour_factor}, it should be noted that $\mathcal{C}$ is not immediately accessible from this state.
In particular, the state contains a sum over $a$ but each term $T^a_{ij}T^a_{jk}$ multiplies a distinct state $\ket{a}_g$ of the gluon register, which means that the desired summation $\sum_a T^a_{ij}T^a_{jk}$ has not yet been performed.

In order to perform the desired sum, we can first observe by inverting eq.~(\ref{eq:Rg_effect}) that if a circuit is constructed for $R_g^{-1}$ and is applied to any state $\sum_{a=1}^8 c_a \ket{a}_g$ of the gluon register, it would produce the state
\begin{equation}
R_g^{-1} \sum_{a=1}^8 c_a \ket{a}_g = \left(\frac{1}{\sqrt{8}} \sum_{a=1}^8 c_a\right)\ket{\Omega}_g + \left(\textrm{terms orthogonal to } \ket{\Omega}_g \right) ,
\end{equation}
which effectively averages over the coefficients of the 8 colour basis states $\ket{a}_g$.
Similarly, it can be seen by inverting eq.~(\ref{eq:Rq_effect}) that a gate $R_q^{-1}$ acting on any state $\sum_{i,k\in\{1,2,3\}} c_{ik} \ket{i}_q \ket{k}_{\tilde{q}}$ of the $q$ and $\tilde{q}$ registers would produce the state
\begin{equation}
R_q^{-1} \sum_{i,k\in\{1,2,3\}} c_{ik} \ket{i}_q \ket{k}_{\tilde{q}} = \left(\frac{1}{\sqrt{3}} \sum_{i=1}^3 c_{ii} \right)\ket{\Omega}_q \ket{\Omega}_{\tilde{q}} + \left(\textrm{terms orthogonal to } \ket{\Omega}_q\ket{\Omega}_{\tilde{q}} \right) ,
\end{equation}
effectively thereby performing a trace over quark colours.
Note that tracing over external colours is not essential, but we have chosen to do so for simplicity as it allows each Feynman diagram to be validated by comparing a single number to the output of our quantum circuits.

Thus, after we have applied the gates $R_g^{-1}$ and $R_q^{-1}$ to the state that was produced in eq.~\eqref{eq:example_circuit_after_second_Q_gate}, we obtain the following state:
\begin{equation}\label{eq:example_trace_final_state}
\frac{1}{24}\left(
\sum_{
\substack{
a \in \{1, ..., 8\}\\
i,j \in \{1,2,3\}
}
} T^a_{ij} T^a_{ji}
\right)
\ket{\Omega}_g\ket{\Omega}_q\ket{\Omega}_{\tilde{q}}\ket{\Omega}_\mathcal{U} + \left(\textrm{terms orthogonal to } \ket{\Omega}_g\ket{\Omega}_q\ket{\Omega}_{\tilde{q}}\ket{\Omega}_\mathcal{U} \right).
\end{equation}
It can be seen that in this state, the coefficient of the original reference state $\ket{\Omega}_g\ket{\Omega}_q\ket{\Omega}_{\tilde{q}}\ket{\Omega}_\mathcal{U}$ encodes the colour factor~\eqref{eq:self_energy_colour_factor} of the diagram.
The factor $\frac{1}{24}$ in eq.~\eqref{eq:example_trace_final_state} is a normalisation that depends on the number of quarks and gluons but does not depend on how they are connected in a particular Feynman diagram.
In the next section we will explain how the example from this section can be generalised to arbitrarily more complicated diagrams by adding more qubits and more $Q$ and $G$ gates.

\section{Calculating the colour factor of arbitrary Feynman diagrams}\label{sec:generalisation}
The illustrative example from sec.~\ref{sec:illustrative_example} can be generalised to calculate colour factors for Feynman diagrams with arbitrary numbers of quarks and gluons.
Given an arbitrary Feynman diagram with $N_q$ quark lines and $N_g$ gluons, the procedure is as follows:
\begin{enumerate}
\item Create a quantum circuit with a 3-qubit gluon register $g$ for each gluon, a pair of 2-qubit quark registers $q$, $\tilde{q}$ for each quark line, and a single unitarisation register $\mathcal{U}$.
\item Initialise each register $r$ to a reference state $\ket{\Omega}_r$ in which each qubit is in the state $\ket{0}$.
\item For each gluon, apply $R_g$ to the corresponding register $g$.
\item For each quark line, apply $R_q$ to the corresponding pair of registers $q,\tilde{q}$.
\item For each quark-gluon vertex, apply a $Q$ gate to the corresponding registers $g$ and $q$.
\item For each triple-gluon vertex, apply a $G$ gate to the 3 corresponding $g$ registers.
\item For each gluon, apply $R_g^{-1}$ to the corresponding gluon register.
\item For each quark, apply $R_q^{-1}$ to the corresponding pair of quark registers $q,\tilde{q}$.
\end{enumerate}
As was the case in the illustrative example in sec.~\ref{sec:illustrative_example}, the colour factor $\mathcal{C}$ for the diagram is now found encoded in the final state of the quantum computer, which is
\begin{equation}\label{eq:final_state}
\frac{1}{\mathcal{N}} \mathcal{C} \ket{\Omega}_{all} + \left(\textrm{terms orthogonal to} \ket{\Omega}_{all}\right) ,
\end{equation}
where $\mathcal{N} = N_c^{n_q} \left(N_c^2-1\right)^{n_g}$ is an overall normalisation and
\begin{equation}
\ket{\Omega}_{all} = \left( \prod_{m=1}^{n_g} \ket{\Omega}_{g_m} \right) \left( \prod_{l=1}^{n_q} \ket{\Omega}_{q_l}\ket{\Omega}_{\tilde{q}_l} \right) \ket{\Omega}_{\mathcal{U}}.
\end{equation}

\section{Validation}\label{sec:validation}
\begin{table}[]
  \caption{\label{tab:checks}
    Colour factors for example Feynman diagrams.
    The first column depicts the Feynman diagrams, with indices on external legs indicating identical colours.
    The central column states the analytical result for the colour factor.
    The last column displays the numerical result for each colour factor obtained using quantum simulations as explained in sec.~\ref{sec:validation} of the text.
    }
  \begin{center}
    \begin{tabular}{c|c|c}
    Diagram & Analytical & Numerical \\
    \midrule
    \thead{\includegraphics[width=0.3\textwidth]{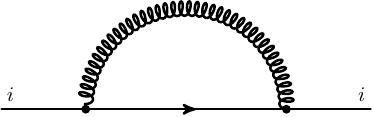}} & $C_F N = 4$ & $3.9988 \pm 0.0012$ %2776113/100000000 Events
    \\
    \midrule
    \thead{\includegraphics[width=0.3\textwidth]{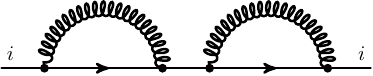}} & ${C_F}^2 N = \frac{16}{3}$ & $5.331 \pm 0.010$ %77082/100000000 Events
    \\
    \midrule
    \thead{\includegraphics[width=0.3\textwidth]{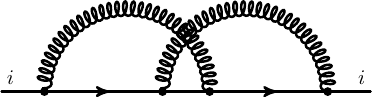}} & $\frac{C_F}{2} = \frac23$ & $0.673 \pm 0.010$ %1229/100000000 Events
    \\
    \midrule
    \thead{\includegraphics[width=0.3\textwidth]{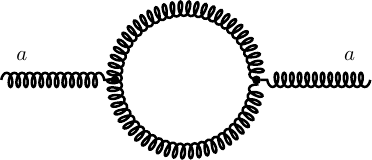}} & $N(N^2-1) = 24$ & $23.95 \pm 0.03$ %218753/100000000 Events
    \\
    \midrule
    \thead{\includegraphics[width=0.3\textwidth]{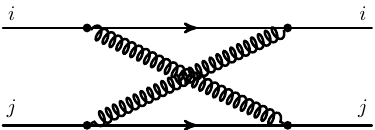}} & $\frac{(N^2-1)}{4} = 2$ & $2.00 \pm 0.03$ %1206/100000000 Events
    \\
    \midrule
    \thead{\includegraphics[width=0.3\textwidth]{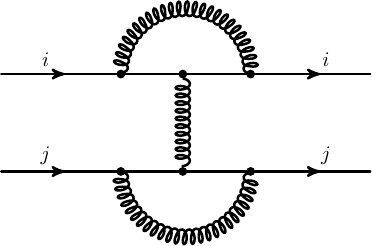}} & $0$ & $0.0^{+0.5}_{-0.0}$ %0/100000000 Events
    \\
    \midrule
    \thead{\includegraphics[width=0.3\textwidth]{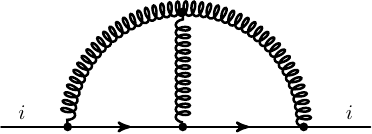}} & $\frac{C_F N^2}{2} = 6$ & $5.92 \pm 0.08$ %1486/100000000 Events
    \\
    \end{tabular}
  \end{center}
\end{table}
The methods of this work were validated by implementing the circuits using the IBM \textsc{Qiskit}~\cite{Qiskit} framework. 
Various circuits were then constructed, each corresponding to a different example of a Feynman diagram shown in Table~\ref{tab:checks}.
These circuits were run on a noiseless simulated quantum computer, using up to 30 simulated qubits.
Each circuit was run $10^8$ times and the output was measured and used to infer the colour factor $\mathcal{C}$ from the fraction of times that the output state was measured to be $\ket{\Omega}_{all}$. 
While this is a straight-forward means to verify that the circuits are functioning correctly, we emphasise that the number of runs required to achieve a given precision can be significantly improved by suitably modifying the circuits and measurement schemes.
Indeed, ongoing follow-up work, which will be reported in a future publication, indicates that it may be straight-forward to achieve the same precision for $\mathcal{C}$ using quadratically fewer runs.
Further improvements may also be possible using schemes such as quantum amplitude estimation~\cite{Brassard:2000,Grinko:2019,Suzuki:2019,Nakaji:2020}, which offers a quadratic speed-up over naive measurements.

\section{Summary and Outlook}
In these conference proceedings, which are based on our article~\cite{Chawdhry:2023jks} and related proceedings~\cite{Chawdhry:2023ken}, we have presented quantum computing circuits to simulate the colour part of perturbative QCD.
As an example application, we showed how these circuits can be used to calculate the colour factors for arbitrary Feynman diagrams.
We implemented the circuits on a simulated noiseless quantum computer and verified them against the analytic expectation for the colour factors of various examples of Feynman diagrams.
This is a first step towards a full quantum simulation of generic perturbative QCD processes.

The work has several avenues for follow-up work.
Firstly, we are now exploring ways to use these circuits to calculate quantum interferences of multiple Feynman diagrams.
Secondly, we can build on this work by also implementing simulations of the kinematic parts of perturbative QCD.
Ultimately, we intend this work to be used to develop a quantum-accelerated Monte-Carlo calculation of scattering rates and cross-sections, obtaining a quadratic speed-up over classical Monte Carlo simulations.

\section*{Acknowledgements}
The authors are grateful to Fabrizio Caola, Stefano Gogioso, Michele Grossi, and Joseph Tooby-Smith for helpful discussions.
The work of H.C.\ was funded by the European Research Council (ERC) under the European Union’s Horizon 2020 research and innovation programme (grant agreement no. 804394 \textsc{hipQCD}) and by the U.S. Department of Energy under grant DE-SC0010102.
M.P.\ acknowledges support by the German Research Foundation (DFG) through the Research Training Group RTG2044.
H.C.\ is grateful to the Galileo Galilei Institute for hospitality and support during the scientific program on ``Theory Challenges in the Precision Era of the Large Hadron Collider,'' where part of this work was performed.

\bibliographystyle{utphys.bst}
\bibliography{ACAT_proceedings_qcolour}

\providecommand{\href}[2]{#2}\begingroup\raggedright\begin{thebibliography}{10}

\bibitem{Chawdhry:2023jks}
H.~A. Chawdhry and M.~Pellen, {\em {Quantum simulation of colour in
  perturbative quantum chromodynamics}}.
  \href{http://dx.doi.org/10.21468/SciPostPhys.15.5.205}{SciPost Phys. {\bf 15}
  (2023) no.~5, 205}, \href{http://arxiv.org/abs/2303.04818}{{\tt
  arXiv:2303.04818 [hep-ph]}}.

\bibitem{Chawdhry:2023ken}
H.~A. Chawdhry and M.~Pellen, {\em {Quantum algorithms for the simulation of
  perturbative QCD processes}}.
  \href{http://dx.doi.org/10.22323/1.432.0087}{PoS {\bf RADCOR2023} (2023)
  087}, \href{http://arxiv.org/abs/2309.06182}{{\tt arXiv:2309.06182
  [hep-ph]}}.

\bibitem{Benioff:1979ce}
P.~Benioff, ``{The computer as a physical system: a microscopic quantum
  mechanical Hamiltonian model of computers as represented by Turing
  machines},'' CPT-79/P-1082.
\newblock 2, 1979.

\bibitem{Feynman:1981tf}
R.~P. Feynman, {\em {Simulating physics with computers}}.
  \href{http://dx.doi.org/10.1007/BF02650179}{Int. J. Theor. Phys. {\bf 21}
  (1982)  467--488}.

\bibitem{Shor.365700}
P.~Shor, \href{http://dx.doi.org/10.1109/SFCS.1994.365700}{``Algorithms for
  quantum computation: discrete logarithms and factoring,''} in {\em
  Proceedings 35th Annual Symposium on Foundations of Computer Science},
  pp.~124--134.
\newblock 1994.

\bibitem{10.1145/237814.237866}
L.~K. Grover, \href{http://dx.doi.org/10.1145/237814.237866}{``A fast quantum
  mechanical algorithm for database search,''} in {\em Proceedings of the
  Twenty-Eighth Annual ACM Symposium on Theory of Computing}, STOC 1996,
  p.~212–219.
\newblock 1996.

\bibitem{ChemistryRev}
Y.~Cao, J.~Romero, J.~P. Olson, M.~Degroote, P.~D. Johnson, M.~Kieferová,
  I.~D. Kivlichan, T.~Menke, B.~Peropadre, N.~P.~D. Sawaya, S.~Sim, L.~Veis,
  and A.~Aspuru-Guzik, {\em Quantum Chemistry in the Age of Quantum Computing}.
  \href{http://dx.doi.org/10.1021/acs.chemrev.8b00803}{Chemical Reviews {\bf
  119} (2019) no.~19, 10856--10915}.

\bibitem{RevModPhys92015003}
S.~McArdle, S.~Endo, A.~Aspuru-Guzik, S.~C. Benjamin, and X.~Yuan, {\em Quantum
  computational chemistry}.
  \href{http://dx.doi.org/10.1103/RevModPhys.92.015003}{Rev. Mod. Phys. {\bf
  92} (2020)  015003}.

\bibitem{RevModPhys.86.153}
I.~M. Georgescu, S.~Ashhab, and F.~Nori, {\em Quantum simulation}.
  \href{http://dx.doi.org/10.1103/RevModPhys.86.153}{Rev. Mod. Phys. {\bf 86}
  (2014)  153--185}.

\bibitem{qute.201900052}
F.~Tacchino, A.~Chiesa, S.~Carretta, and D.~Gerace, {\em Quantum Computers as
  Universal Quantum Simulators: State-of-the-Art and Perspectives}.
  \href{http://dx.doi.org/https://doi.org/10.1002/qute.201900052}{Advanced
  Quantum Technologies {\bf 3} (2020) no.~3, 1900052}.

\bibitem{Klco:2021lap}
N.~Klco, A.~Roggero, and M.~J. Savage, {\em {Standard model physics and the
  digital quantum revolution: thoughts about the interface}}.
  \href{http://dx.doi.org/10.1088/1361-6633/ac58a4}{Rept. Prog. Phys. {\bf 85}
  (2022) no.~6, 064301}, \href{http://arxiv.org/abs/2107.04769}{{\tt
  arXiv:2107.04769 [quant-ph]}}.

\bibitem{Bauer:2022hpo}
C.~W. Bauer {\em et al.}, {\em {Quantum Simulation for High-Energy Physics}}.
  \href{http://dx.doi.org/10.1103/PRXQuantum.4.027001}{PRX Quantum {\bf 4}
  (2023) no.~2, 027001}, \href{http://arxiv.org/abs/2204.03381}{{\tt
  arXiv:2204.03381 [quant-ph]}}.

\bibitem{Bepari:2020xqi}
K.~Bepari, S.~Malik, M.~Spannowsky, and S.~Williams, {\em {Towards a quantum
  computing algorithm for helicity amplitudes and parton showers}}.
  \href{http://dx.doi.org/10.1103/PhysRevD.103.076020}{Phys. Rev. D {\bf 103}
  (2021) no.~7, 076020}, \href{http://arxiv.org/abs/2010.00046}{{\tt
  arXiv:2010.00046 [hep-ph]}}.

\bibitem{Bauer:2019qxa}
C.~W. Bauer, W.~A. de~Jong, B.~Nachman, and D.~Provasoli, {\em {Quantum
  Algorithm for High Energy Physics Simulations}}.
  \href{http://dx.doi.org/10.1103/PhysRevLett.126.062001}{Phys. Rev. Lett. {\bf
  126} (2021) no.~6, 062001}, \href{http://arxiv.org/abs/1904.03196}{{\tt
  arXiv:1904.03196 [hep-ph]}}.

\bibitem{Bepari:2021kwv}
K.~Bepari, S.~Malik, M.~Spannowsky, and S.~Williams, {\em {Quantum walk
  approach to simulating parton showers}}.
  \href{http://dx.doi.org/10.1103/PhysRevD.106.056002}{Phys. Rev. D {\bf 106}
  (2022) no.~5, 056002}, \href{http://arxiv.org/abs/2109.13975}{{\tt
  arXiv:2109.13975 [hep-ph]}}.

\bibitem{Gustafson:2022dsq}
G.~Gustafson, S.~Prestel, M.~Spannowsky, and S.~Williams, {\em {Collider events
  on a quantum computer}}.
  \href{http://dx.doi.org/10.1007/JHEP11(2022)035}{JHEP {\bf 11} (2022)  035},
  \href{http://arxiv.org/abs/2207.10694}{{\tt arXiv:2207.10694 [hep-ph]}}.

\bibitem{Brassard:2000}
G.~Brassard, P.~H{\o}yer, M.~Mosca, and A.~Tapp, {\em {Quantum Amplitude
  Amplification and Estimation}}.
  \href{http://dx.doi.org/10.1090/conm/305/05215}{Quantum Computation and
  Information {\bf 305} (2002)  },
  \href{http://arxiv.org/abs/quant-ph/0005055}{{\tt arXiv:quant-ph/0005055
  [quant-ph]}}.

\bibitem{Grinko:2019}
D.~Grinko, J.~Gacon, C.~Zoufal, and S.~Woerner, {\em {Iterative Quantum
  Amplitude Estimation}}.
  \href{http://dx.doi.org/10.1038/s41534-021-00379-1}{npj Quantum Inf {\bf 7}
  (2021) no.~52, }, \href{http://arxiv.org/abs/1912.05559}{{\tt
  arXiv:1912.05559 [quant-ph]}}.

\bibitem{Suzuki:2019}
Y.~Suzuki, S.~Uno, R.~Raymond, T.~Tanaka, T.~Onodera, and N.~Yamamoto, {\em
  {Amplitude estimation without phase estimation}}.
  \href{http://dx.doi.org/10.1007/s11128-019-2565-2}{Quantum Information
  Processing {\bf 19} (2020) no.~75, },
  \href{http://arxiv.org/abs/1904.10246}{{\tt arXiv:1904.10246 [quant-ph]}}.

\bibitem{Nakaji:2020}
K.~Nakaji, {\em {Faster Amplitude Estimation}}.
  \href{http://dx.doi.org/10.26421/QIC20.13-14-2}{Quantum Information \&
  Computation 2020 {\bf 20} (2020) no.~13\&14, },
  \href{http://arxiv.org/abs/2003.02417}{{\tt arXiv:2003.02417 [quant-ph]}}.

\bibitem{Agliardi:2022ghn}
G.~Agliardi, M.~Grossi, M.~Pellen, and E.~Prati, {\em {Quantum integration of
  elementary particle processes}}.
  \href{http://dx.doi.org/10.1016/j.physletb.2022.137228}{Phys. Lett. B {\bf
  832} (2022)  137228}, \href{http://arxiv.org/abs/2201.01547}{{\tt
  arXiv:2201.01547 [hep-ph]}}.

\bibitem{Bauer:2021gup}
C.~W. Bauer, M.~Freytsis, and B.~Nachman, {\em {Simulating Collider Physics on
  Quantum Computers Using Effective Field Theories}}.
  \href{http://dx.doi.org/10.1103/PhysRevLett.127.212001}{Phys. Rev. Lett. {\bf
  127} (2021) no.~21, 212001}, \href{http://arxiv.org/abs/2102.05044}{{\tt
  arXiv:2102.05044 [hep-ph]}}.

\bibitem{Bravo-Prieto:2021ehz}
C.~Bravo-Prieto, J.~Baglio, M.~C\`e, A.~Francis, D.~M. Grabowska, and
  S.~Carrazza, {\em {Style-based quantum generative adversarial networks for
  Monte Carlo events}}.
  \href{http://dx.doi.org/10.22331/q-2022-08-17-777}{Quantum {\bf 6} (2022)
  777}, \href{http://arxiv.org/abs/2110.06933}{{\tt arXiv:2110.06933
  [quant-ph]}}.

\bibitem{Cervera-Lierta:2017tdt}
A.~Cervera-Lierta, J.~I. Latorre, J.~Rojo, and L.~Rottoli, {\em {Maximal
  Entanglement in High Energy Physics}}.
  \href{http://dx.doi.org/10.21468/SciPostPhys.3.5.036}{SciPost Phys. {\bf 3}
  (2017) no.~5, 036}, \href{http://arxiv.org/abs/1703.02989}{{\tt
  arXiv:1703.02989 [hep-th]}}.

\bibitem{Clemente:2022nll}
G.~Clemente, A.~Crippa, K.~Jansen, S.~Ram\'\i{}rez-Uribe, A.~E.
  Renter\'\i{}a-Olivo, G.~Rodrigo, G.~F.~R. Sborlini, and L.~Vale~Silva, {\em
  {Variational quantum eigensolver for causal loop Feynman diagrams and
  directed acyclic graphs}}.
  \href{http://dx.doi.org/10.1103/PhysRevD.108.096035}{Phys. Rev. D {\bf 108}
  (2023) no.~9, 096035}, \href{http://arxiv.org/abs/2210.13240}{{\tt
  arXiv:2210.13240 [hep-ph]}}.

\bibitem{Cruz-Martinez:2023vgs}
J.~M. Cruz-Martinez, M.~Robbiati, and S.~Carrazza, {\em {Multi-variable
  integration with a variational quantum circuit}}.
  \href{http://dx.doi.org/10.1088/2058-9565/ad5866}{Quantum Sci. Technol. {\bf
  9} (2024) no.~3, 035053}, \href{http://arxiv.org/abs/2308.05657}{{\tt
  arXiv:2308.05657 [quant-ph]}}.

\bibitem{Fedida:2022izl}
S.~Fedida and A.~Serafini, {\em {Tree-level entanglement in quantum
  electrodynamics}}. \href{http://dx.doi.org/10.1103/PhysRevD.107.116007}{Phys.
  Rev. D {\bf 107} (2023) no.~11, 116007},
  \href{http://arxiv.org/abs/2209.01405}{{\tt arXiv:2209.01405 [quant-ph]}}.

\bibitem{Kiss:2022pjw}
O.~Kiss, M.~Grossi, E.~Kajomovitz, and S.~Vallecorsa, {\em {Conditional Born
  machine for Monte Carlo event generation}}.
  \href{http://dx.doi.org/10.1103/PhysRevA.106.022612}{Phys. Rev. A {\bf 106}
  (2022) no.~2, 022612}, \href{http://arxiv.org/abs/2205.07674}{{\tt
  arXiv:2205.07674 [quant-ph]}}.

\bibitem{Li:2021kcs}
{\bf QuNu} Collaboration, T.~Li, X.~Guo, W.~K. Lai, X.~Liu, E.~Wang, H.~Xing,
  D.-B. Zhang, and S.-L. Zhu, {\em {Partonic collinear structure by quantum
  computing}}. \href{http://dx.doi.org/10.1103/PhysRevD.105.L111502}{Phys. Rev.
  D {\bf 105} (2022) no.~11, L111502},
  \href{http://arxiv.org/abs/2106.03865}{{\tt arXiv:2106.03865 [hep-ph]}}.

\bibitem{Perez-Salinas:2020nem}
A.~P\'erez-Salinas, J.~Cruz-Martinez, A.~A. Alhajri, and S.~Carrazza, {\em
  {Determining the proton content with a quantum computer}}.
  \href{http://dx.doi.org/10.1103/PhysRevD.103.034027}{Phys. Rev. D {\bf 103}
  (2021) no.~3, 034027}, \href{http://arxiv.org/abs/2011.13934}{{\tt
  arXiv:2011.13934 [hep-ph]}}.

\bibitem{Ramirez-Uribe:2021ubp}
S.~Ram\'\i{}rez-Uribe, A.~E. Renter\'\i{}a-Olivo, G.~Rodrigo, G.~F.~R.
  Sborlini, and L.~Vale~Silva, {\em {Quantum algorithm for Feynman loop
  integrals}}. \href{http://dx.doi.org/10.1007/JHEP05(2022)100}{JHEP {\bf 05}
  (2022)  100}, \href{http://arxiv.org/abs/2105.08703}{{\tt arXiv:2105.08703
  [hep-ph]}}.

\bibitem{Rigobello:2023ype}
M.~Rigobello, G.~Magnifico, P.~Silvi, and S.~Montangero, {\em {Hadrons in
  (1+1)D Hamiltonian hardcore lattice QCD}}.
  \href{http://arxiv.org/abs/2308.04488}{{\tt arXiv:2308.04488 [hep-lat]}}.

\bibitem{Williams:2023muq}
S.~J. Williams, \href{http://dx.doi.org/10.25560/105867}{{\em {Event generation
  on quantum computers}}}.
\newblock PhD thesis, Imperial Coll., London, 2023.

\bibitem{Nicotra:2023rmn}
D.~Nicotra, M.~Lucio~Martinez, J.~A. de~Vries, M.~Merk, K.~Driessens, R.~L.
  Westra, D.~Dibenedetto, and D.~H. C\'ampora~P\'erez, {\em {A quantum
  algorithm for track reconstruction in the LHCb vertex detector}}.
  \href{http://dx.doi.org/10.1088/1748-0221/18/11/P11028}{JINST {\bf 18} (2023)
  no.~11, P11028}, \href{http://arxiv.org/abs/2308.00619}{{\tt arXiv:2308.00619
  [quant-ph]}}.

\bibitem{Nagano:2023kge}
L.~Nagano, A.~Miessen, T.~Onodera, I.~Tavernelli, F.~Tacchino, and K.~Terashi,
  {\em {Quantum data learning for quantum simulations in high-energy physics}}.
  \href{http://dx.doi.org/10.1103/PhysRevResearch.5.043250}{Phys. Rev. Res.
  {\bf 5} (2023) no.~4, 043250}, \href{http://arxiv.org/abs/2306.17214}{{\tt
  arXiv:2306.17214 [quant-ph]}}.

\bibitem{Turco:2023rmx}
M.~Turco, G.~M. Quinta, J.~a. Seixas, and Y.~Omar, {\em {Quantum Simulation of
  Bound State Scattering}}.
  \href{http://dx.doi.org/10.1103/PRXQuantum.5.020311}{PRX Quantum {\bf 5}
  (2024) no.~2, 020311}, \href{http://arxiv.org/abs/2305.07692}{{\tt
  arXiv:2305.07692 [quant-ph]}}.

\bibitem{Bass:2023hoi}
S.~D. Bass and M.~Doser, {\em {Quantum sensing for particle physics}}.
  \href{http://dx.doi.org/10.1038/s42254-024-00714-3}{Nature Rev. Phys. {\bf 6}
  (2024) no.~5, 329--339}, \href{http://arxiv.org/abs/2305.11518}{{\tt
  arXiv:2305.11518 [quant-ph]}}.

\bibitem{DiMeglio:2023nsa}
A.~Di~Meglio {\em et al.}, {\em {Quantum Computing for High-Energy Physics:
  State of the Art and Challenges. Summary of the QC4HEP Working Group}}.
  \href{http://arxiv.org/abs/2307.03236}{{\tt arXiv:2307.03236 [quant-ph]}}.

\bibitem{Bermot:2023kvh}
E.~Bermot, C.~Zoufal, M.~Grossi, J.~Schuhmacher, F.~Tacchino, S.~Vallecorsa,
  and I.~Tavernelli, {\em {Quantum Generative Adversarial Networks For Anomaly
  Detection In High Energy Physics}}.
  \href{http://arxiv.org/abs/2304.14439}{{\tt arXiv:2304.14439 [quant-ph]}}.

\bibitem{Sborlini:2023uyq}
G.~F.~R. Sborlini, {\em {Geometrical causality: casting Feynman integrals into
  quantum algorithms}}.
  \href{http://dx.doi.org/10.31349/SuplRevMexFis.4.021103}{Rev. Mex. Fis.
  Suppl. {\bf 4} (2023) no.~2, 021103},
  \href{http://arxiv.org/abs/2305.08550}{{\tt arXiv:2305.08550 [hep-ph]}}.

\bibitem{Humble:2022klb}
T.~S. Humble, G.~N. Perdue, and M.~J. Savage, {\em {Snowmass Computational
  Frontier: Topical Group Report on Quantum Computing}}.
  \href{http://arxiv.org/abs/2209.06786}{{\tt arXiv:2209.06786 [quant-ph]}}.

\bibitem{Hayata:2023bgh}
T.~Hayata and Y.~Hidaka, {\em {q deformed formulation of Hamiltonian SU(3)
  Yang-Mills theory}}. \href{http://dx.doi.org/10.1007/JHEP09(2023)123}{JHEP
  {\bf 09} (2023)  123}, \href{http://arxiv.org/abs/2306.12324}{{\tt
  arXiv:2306.12324 [hep-lat]}}.

\bibitem{Sborlini:2023mws}
G.~F.~R. Sborlini, {\em {Tackling Feynman integrals with quantum minimization
  algorithms}}. \href{http://dx.doi.org/10.22323/1.449.0501}{PoS {\bf
  EPS-HEP2023} (2024)  501}, \href{http://arxiv.org/abs/2309.12739}{{\tt
  arXiv:2309.12739 [hep-th]}}.

\bibitem{Brown:2023llg}
C.~Brown, M.~Spannowsky, A.~Tapper, S.~Williams, and I.~Xiotidis, {\em {Quantum
  pathways for charged track finding in high-energy collisions}}.
  \href{http://dx.doi.org/10.3389/frai.2024.1339785}{Front. Artif. Intell. {\bf
  7} (2024)  1339785}, \href{http://arxiv.org/abs/2311.00766}{{\tt
  arXiv:2311.00766 [hep-ph]}}.

\bibitem{Rodrigo:2024say}
G.~Rodrigo, {\em {Quantum Algorithms in Particle Physics}}.
  \href{http://dx.doi.org/10.5506/APhysPolBSupp.17.2-A14}{Acta Phys. Polon.
  Supp. {\bf 17} (2024) no.~2, 2--A14},
  \href{http://arxiv.org/abs/2401.16208}{{\tt arXiv:2401.16208 [hep-ph]}}.

\bibitem{Ramirez-Uribe:2024wua}
S.~Ram\'\i{}rez-Uribe, A.~E. Renter\'\i{}a-Olivo, and G.~Rodrigo, {\em {Quantum
  querying based on multicontrolled Toffoli gates for causal Feynman loop
  configurations and directed acyclic graphs}}.
  \href{http://arxiv.org/abs/2404.03544}{{\tt arXiv:2404.03544 [quant-ph]}}.

\bibitem{Galvez-Viruet:2024hry}
J.~J. G\'alvez-Viruet and F.~J. Llanes-Estrada, {\em {A dynamical
  implementation of canonical second quantization on a quantum computer}}.
  \href{http://arxiv.org/abs/2406.03147}{{\tt arXiv:2406.03147 [hep-th]}}.

\bibitem{Qiskit}
S.~Anis {\em et al.}, \href{http://dx.doi.org/10.5281/zenodo.2573505}{{\em
  Qiskit: An Open-source Framework for Quantum Computing}}, 2021.

\end{thebibliography}\endgroup

\end{document}